\documentclass[secnumarabic,amssymb, amsmath,
nofootinbib,tightenlines, nobibnotes, aps, prl]{revtex4}
\usepackage{graphicx}
\begin{document}

\title{ CP violation in Semi-Leptonic $\tau$ decays}


\author{David Delepine}
\email{delepine@fisica.ugto.mx}
\affiliation{Instituto de F\'{\i}sica, Universidad de Guanajuato \\
Loma del Bosque \# 103, Lomas del Campestre, \\
37150 Le\'on, Guanajuato; M\'exico}


\begin{abstract}
We study CP violation in semi-leptonic $\tau$ decays and we determine the conditions necessary to be able to define a observable CP asymmetry. We apply these conditions in both models, the standard model for the electroweak interactions and its supersymmetric extensions. In the first case, the leading order contribution to the direct CP asymmetry in
$\tau^{\pm} \to K^{\pm}\pi^0 \nu_{\tau}$ decay rates is evaluated. In the second case,we compute the SUSY effective hamiltonian
that describes the $|\Delta S|=1$ semileptonic decays of tau
leptons. We show that SUSY contributions may
enhance the CP asymmetry of $\tau \to K \pi \nu_{\tau}$  decays by
several orders of magnitude compared to the standard model expectations.
\end{abstract}

\date{\today}

\maketitle

\section{Introduction}
Experimental searches for CP violating asymmetries in tau
lepton semileptonic decays have been carried out in the $\tau \to \pi\pi
\nu_{\tau}$  \cite{pipi} and  $\tau \to K_s  \pi \nu_{\tau}$ \cite{cleo} modes.
 Motivation for these searches in the context of beyond the Standard Model
approaches were provided in refs. \cite{cptau,tsai}. In ref. \cite{cleo},
the missing evidence for a non-zero CP asymmetry was interpreted in terms
of a (CP-violating) coupling $\Lambda$ due to a
charged scalar exchange and the limit $-0.172 <Im (\Lambda)<0.067$ (at
90\% c.l.) has been derived. The CP-odd observable studied in \cite{cleo}
depends upon two variables of a particular kinematical distribution
of semileptonic tau  decays  as long as  this  effect is assumed to have
its origin in the interference of  scalar and  vector form factors
In this talk, we shall first determine the general conditions to get observable CP asymmetry. Then we should apply it to
Standard Model and its supersymmetric extensions

 The general amplitude for $\tau^-(p) \rightarrow
K^-(k)\pi^0(k')\nu_{\tau}(p')$is
given by
 \begin{eqnarray}
  {\cal M} &=&
\frac{G_FV_{us}}{\sqrt{2}}
\left\{\bar{u}(p')\gamma^{\mu}(1-\gamma_5)u(p) F_V(t)
\left[(k-k')_{\mu}-\frac{\Delta^2}{t}q_{\mu} \right] \right. \nonumber \\
& & + \bar{u}(p')(1+\gamma_5)u(p) m_{\tau} \Lambda
F_S(t)\frac{\Delta^2}{t} \nonumber
\\
& & \left. + F_{T}\langle K\pi |\bar{s}\sigma _{\mu \upsilon
}u|0\rangle \bar{u}(p^{\prime })\sigma ^{\mu \upsilon }(1+\gamma_5)
u(p)\right\}, \nonumber \end{eqnarray}

where $q=k+k'$ ($t=q^2$) is the momentum transfer
to the hadronic system, $\Delta^2   \equiv
m_K^2-m_{\pi}^2$ and $F_{V,S,T}(t)$ are the {\it effective} form
factors describing the hadronic matrix elements ({In
Standard Model, $F_T=0$, $\Lambda=1$)

\begin{eqnarray}
\sum_{pols}\vert\mathcal{M}\vert^2&\sim &\vert
F_V\vert^2(2p.Qp^{\prime}.Q-p.p^{\prime}Q^2)+\vert\Lambda \vert^2
\vert
F_S(t)\vert^2M^2p.p^{\prime}  \nonumber \\
&& +2Re\Lambda\cdot Re(F_SF_V^*)M m_{\tau}p^{\prime}.Q
-2Im\Lambda\cdot Im(F_SF_V^*)Mm_{\tau}p^{\prime}.Q \nonumber
\end{eqnarray}
where $Q_{\mu}=(k-k')_{\mu}-\frac{\Delta^2}{t}q_{\mu}$and $F_T$ contributions have been neglected. The last term is odd under a CP transformation but
the last two terms disappear once we integrate on
the kinematical variable $u$.

It is not possible to generate a CP asymmetry in total decay rates
corresponding to this process unless:

\begin{itemize}
  \item $F_{V,S}=f_{V,S}+ a_{V,S}$ which $a_{V,S}$
   containing a weak CP violating phase and a different strong phase than
   $f_{V,S}$
  \item Interference between tensorial and vector contribution
  $\Rightarrow$ we need to go beyond SM.
  \item to look for
the double differential distribution
$(d^{2}\Gamma /dudt)$ or a variant of it as CLEO collaboration did \cite{cleo}.
\end{itemize}


\section{CP asymmetry in Standard Model}

The form factor $f_{V,S}(t)$ are dominated at tree
level by a single vector or scalar strange resonance:
\begin{equation}
 f_i(t)=\frac{f_i(0)m_i^2}{m_i^2-t-im_i\Gamma_i},
\ \ \ i=V,\ S ,\nonumber
 \end{equation}
 where ($m_i,\ \Gamma_i$) denote the mass and width of the resonance
in the vector or scalar configuration (respectively
the$K^*(892)$ or
 $K^*_0(1430)$).

 $ \Rightarrow$  the tree-level strong phase
fixed by the decay width of these resonances, while
the weak phase is absent at the tree-level. Higher
order contributions can induce weak phase contributions
($a_{V,S}$) (see figure \ref{fig1}) \footnote{for details of the computation, see reference \cite{david1}}.
\begin{figure}
  \includegraphics[width=10cm]{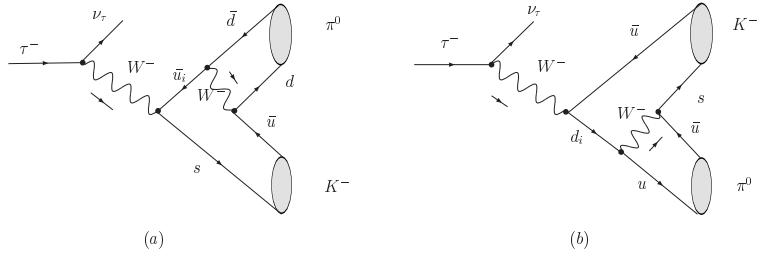}\\
  \caption{Higher order contributions to CP asymmetry in Standard Model}
  \label{fig1}
\end{figure}
\begin{eqnarray}
A_{CP}&=&\frac{\Gamma(\tau^+ \to
K^+\pi^0\bar{\nu}_{\tau})-\Gamma(\tau^- \to
K^-\pi^0\nu_{\tau})}{\Gamma(\tau^+ \to
 K^+\pi^0\bar{\nu}_{\tau})+\Gamma(\tau^-
\to K^-\pi^0\nu_{\tau})} \nonumber \\
&\approx &
-\frac{\sqrt{2}G_F^3m_{\tau}^5\mbox{Im}(V_{us}V_{ud}^*V_{td}V_{ts}^*)f_Kf_{\pi}}{
768 \pi^3\Gamma (\tau^+ \to K^+\pi^0 \bar{\nu}_{\tau})}\times
I_{CP}\ \nonumber,
\end{eqnarray}
where
\begin{equation}
I_{CP}=\frac{1}{m_{\tau}^6}\int_{(m_K+m_{\pi})^2}^{m_{\tau}^2}\frac{dt}{t^3}
(m_{\tau}^2-t)^2 h(t) \left(1+\frac{2t}{m_{\tau}^2}\right)
\lambda^{3/2}(t,m_K^2,m_{\pi}^2)\nonumber
\end{equation}
where
$\lambda(x,y,z)=x^2+y^2+z^2-2xy-2xz-2yz$
\begin{eqnarray} A_{CP}&\approx-
&\frac{\sqrt{2}G_F\mbox{Im}(V_{us}V_{ud}^*V_{td}V_{ts}^*)f_Kf_{\pi}}{20
B(\tau^+ \to K^+\pi^0 \bar{\nu}_{\tau})} \times I_{CP} \approx 2.3
\times 10^{-12},\nonumber
\end{eqnarray}
with $B(\tau^+ \to K^+\pi^0 \bar{\nu}_{\tau}) =(4.5
\pm 0.3) \times 10^{-3}$  and  $m_c= 1.35$ GeV,

\section{CP asymmetry in Supersymmetric Models}
Strangeness-changing $\vert \Delta S \vert=1$ decays of tau leptons are
driven by the $\tau^- \to \bar{u}s\nu_{\tau}$ elementary process.
The effective Hamiltonian $H_{eff}$ derived from SUSY can be expressed as
\begin{equation}
H_{eff}=\frac{G_{F}}{\sqrt{2}}V_{us}\sum_{i}C_{i}(\mu )Q_{i}(\mu
), \label{SHeff}
\end{equation}%
where $C_{i}$ are the Wilson coefficients and $Q_{i}$ are the relevant local
operators at low energy scale $\mu \simeq m_{\tau }$. The operators are
given by
\begin{eqnarray}
Q_{1} &=&(\bar{\nu}\gamma ^{\mu }L\tau )(\bar{s}\gamma _{\mu }Lu), \\
Q_{2} &=&(\bar{\nu}\gamma _{\mu }L\tau )(\bar{s}\gamma _{\mu }Ru), \\
Q_{3} &=&(\bar{\nu}R\tau )(\bar{s}Lu), \\
Q_{4} &=&(\bar{\nu}R\tau )(\bar{s}Ru), \\
Q_{5} &=&(\bar{\nu}\sigma _{\mu \nu }R\tau )(\bar{s}\sigma ^{\mu \nu }Ru).
\end{eqnarray}%
where $L,R$ are as defined in the previous section and $\sigma ^{\mu \nu }=%
\frac{i}{2}[\gamma ^{\mu },\gamma ^{\nu }]$. The Wilson coefficients $C_{i}$%
, at the electroweak scale, can be expressed as $%
C_{i}=C_{i}^{SM}+C_{i}^{SUSY}$ where $C_{i}^{SM}$ are given by
\begin{eqnarray}
&&C_{1}^{SM}=1,  \nonumber \\
&&C_{2,3,4,5}^{SM}=0.
\end{eqnarray}%
SUSY contributions to the Hamiltonian of $\tau ^{-}\rightarrow \bar{u}s\nu
_{\tau }$ transitions can be generated through two topological box diagrams
as shown in Figs. (\ref{SUSYfig},\ref{SUSYfig2}). Other SUSY contributions
(vertex corrections) are suppressed either due to small Yukawa couplings of
light quarks or because they have the same structure as the SM in the
hadronic vertex. 
\begin{figure}
  \includegraphics[width=14cm]{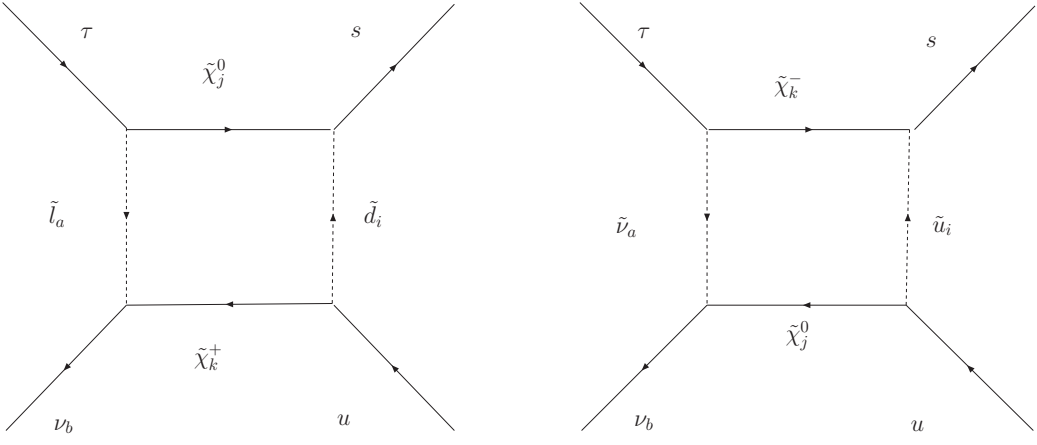}\\
  \caption{SUSY box contributions to $\protect\tau ^{-}\rightarrow \bar{u}s%
\protect\nu _{\protect\tau }$ transition.}
\label{SUSYfig}
  \end{figure}
\begin{figure}
  \includegraphics[width=14cm]{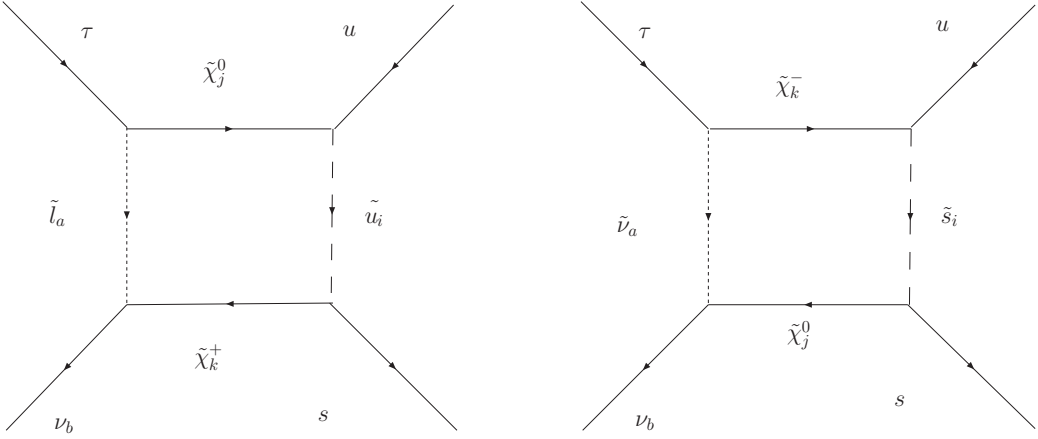}\\
  \caption{Crossed diagrams of Figure 2.}
\label{SUSYfig2}
  \end{figure}

In our computations of Wilson coefficients, we will work in the mass
insertion approximation (MIA), where gluino and neutralino are flavor
diagonal. Denoting by $(\Delta _{AB}^{f})_{ab}$ the \textit{off-}diagonal
terms in the sfermion mass matrices where $A,B$ indicate chirality, $%
A,B=(L,R)$, the $A-B$ sfermion propagator can be expanded as
\begin{equation}
\langle \tilde{f}_{A}^{a}\tilde{f}_{B}^{b\ast }\rangle =i(k^{2}I-\tilde{m}%
^{2}I-\Delta _{AB}^{f}))_{ab}^{-1}\simeq \frac{i\delta _{ab}}{k^{2}-\tilde{m}%
^{2}}+\frac{i(\Delta _{AB}^{f})_{ab}}{(k^{2}-\tilde{m}^{2})^{2}}+O(\Delta
^{2}),
\end{equation}%
where $\tilde{f}$ denotes any scalar fermion, $a,b=(1,2,3)$ are flavor
indices, $I$ is the unit matrix, and $\tilde{m}$ is the average sfermion
mass. It is convenient to define a dimensionless quantity $(\delta
_{AB}^{f})_{ab}\equiv (\Delta _{AB}^{f})_{ab}/\tilde{m}^{2}.$ As long as $%
(\Delta _{AB}^{f})_{ab}$ is smaller than $\tilde{m}^{2}$ we can consider
only the first order term in $(\delta _{AB}^{f})_{ab}$ of the sfermion
propagator expansion. In our analysis we will keep only terms proportional
to the third generation Yukawa couplings and terms of order $\lambda $ where
$\lambda =V_{us}.$

The complete expressions for the Wilson coefficients $C_i$ at
$m_W$ scale induced by SUSY computed from Figs.
(\ref{SUSYfig},\ref{SUSYfig2}) can be found in reference \cite{david2}. As
can be seen in ref.\cite{david2}, the $C_i$ are given in terms of
several mass insertions that represent the flavor transitions
between different generations of quarks or leptons. In general,
these mass insertions are complex and of order one. However, the
experimental limits of several flavor changing neutral currents
impose severe constraints on most of these mass insertions. In the
following, we summarize all the important constraints on the
relevant mass insertions for our process.

\begin{enumerate}
\item From the experimental measurements of $BR(\mu \to e \gamma) < 1.2
\times 10^{-11}$, the following bounds on $\vert(\delta^l_{12})_{AB}\vert$
and $\vert(\delta^{\nu}_{12})_{AB}\vert$ are obtained \cite{Branco:2003zy}:
For $M_1 \sim M_2=100$ GeV and $\mu =\tilde{m}_l=200$ GeV,
\begin{eqnarray}
&&\vert(\delta^l_{12})_{LL}\vert \raise0.3ex\hbox{$\;<$\kern-0.75em%
\raise-1.1ex\hbox{$\sim\;$}} 10^{-3}, ~~~~~~~~
\vert(\delta^l_{12})_{LR}\vert \raise0.3ex\hbox{$\;<$\kern-0.75em%
\raise-1.1ex\hbox{$\sim\;$}} 10^{-6}, \\
&&\vert(\delta^{\nu}_{12})_{LL}\vert \raise0.3ex\hbox{$\;<$\kern-0.75em%
\raise-1.1ex\hbox{$\sim\;$}} 6\times 10^{-4}, ~~~
\vert(\delta^{\nu}_{13})_{LL}\vert \raise0.3ex\hbox{$\;<$\kern-0.75em%
\raise-1.1ex\hbox{$\sim\;$}} 4\times 10^{-4}, ~~~~
\vert(\delta^{\nu}_{23})_{LL}\vert \raise0.3ex\hbox{$\;<$\kern-0.75em%
\raise-1.1ex\hbox{$\sim\;$}} 7\times 10^{-4}.~~~~~~~
\end{eqnarray}

\item From $BR(\tau \to \mu \gamma) < 1.1 \times 10^{-6}$, one gets the
following constraint on $\vert (\delta^l_{23})_{LR})\vert$ \cite%
{Gabbiani:1996hi}:
\begin{equation}
\vert(\delta^l_{23})_{LR}\vert \raise0.3ex\hbox{$\;<$\kern-0.75em%
\raise-1.1ex\hbox{$\sim\;$}} 2 \times 10^{-2},
\end{equation}
and from $BR(\tau \to e \gamma) < 2.7 \times 10^{-6}$, one finds \cite%
{Gabbiani:1996hi}:
\begin{equation}
\vert(\delta^l_{13})_{LR}\vert \raise0.3ex\hbox{$\;<$\kern-0.75em%
\raise-1.1ex\hbox{$\sim\;$}} 1 \times 10^{-1}.
\end{equation}

\item The mass insertions $(\delta_{12}^d)_{AB}$ are constrained by the $%
\Delta M_K $ and $\epsilon_K $ as follows \cite{Gabbiani:1996hi}:
\begin{eqnarray}
\vert(\delta^d_{12})_{LL}\vert \raise0.3ex\hbox{$\;<$\kern-0.75em%
\raise-1.1ex\hbox{$\sim\;$}} 4\times 10^{-2}, ~~~~~~~~~~~~~~~~
\vert(\delta^d_{12})_{LR}\vert \raise0.3ex\hbox{$\;<$\kern-0.75em%
\raise-1.1ex\hbox{$\sim\;$}} 4 \times 10^{-3}, \\
\sqrt{\vert\mathrm{{Im}\left[(\delta^d_{12})_{LL}\right]^2\vert}} \raise0.3ex%
\hbox{$\;<$\kern-0.75em\raise-1.1ex\hbox{$\sim\;$}} 3\times 10^{-3}, ~~~~~~
\sqrt{\vert\mathrm{{Im}\left[(\delta^d_{12})_{LR}\right]^2\vert}} \raise0.3ex%
\hbox{$\;<$\kern-0.75em\raise-1.1ex\hbox{$\sim\;$}} 3\times 10^{-4}.
\end{eqnarray}

\item The mass insertion $(\delta_{12}^u)_{AB}$ are constrained by the $%
\Delta M_D$ as follows \cite{Khalil:2006zb}:
\begin{equation}
\vert(\delta^u_{12})_{LL}\vert \raise0.3ex\hbox{$\;<$\kern-0.75em%
\raise-1.1ex\hbox{$\sim\;$}} 1.7\times 10^{-2}, ~~~~~~~~~~~~~~~
\vert(\delta^u_{12})_{LR}\vert \raise0.3ex\hbox{$\;<$\kern-0.75em%
\raise-1.1ex\hbox{$\sim\;$}} 2.4 \times 10^{-2}.
\end{equation}
\end{enumerate}

Here, three comments are in order. $i)$ Due to the Hermiticity of the $LL$
sector in the sfermion mass matrix, $(\delta _{AB}^{f})_{LL}=(\delta
_{AB}^{f})_{LL}^{\dagger }=(\delta _{BA}^{f})_{LL}^{\ast }$, where $%
A,B=1,2,3 $. $ii)$ The above constraints imposed on the mass insertions $%
(\delta _{AB}^{q,l})_{LL,LR}$ are derived from supersymmetric
contributions through exchange of gluino or neutralino which
preserves chirality, therefore same constraints are also imposed
on the mass insertions $(\delta _{AB}^{q,l})_{RR,RL}$. $iii)$ The
mass insertions $(\delta
_{AB}^{f})_{LR(RL)}$ are not, in general, related to the mass insertions $%
(\delta _{BA}^{f})_{LR(RL)}$. Taking the above constraints into account, one
finds that the dominant contribution to the $\tau ^{-}\rightarrow u\bar{s}%
\nu _{\tau }$ is given in terms of $(\delta _{32}^{\nu })_{LR}$, $(\delta
_{32}^{\nu })_{RL}$, $(\delta _{21}^{d})_{RL}$ and $(\delta _{21}^{u})_{LR}$%
. Notice that the effective Hamiltonian (eq. \ref{SHeff}) derived
in this section can induce supersymmetric effects in all the
$|\Delta S|=1$ exclusive $\tau $ lepton decay.

The total amplitude (SM and SUSY) of the
$\tau(p)\rightarrow K(q)\pi (q^{\prime })\nu _{\tau
}(p^{\prime })$decay as%

\begin{eqnarray} %
\mathcal{A}_{T}(\tau \rightarrow K\pi \nu ) &=&\frac{G_{F}V_{us}}{\sqrt{2}}%
\Big[ (1+C_{1})\langle K\pi |\bar{s}\gamma _{\mu }u|0\rangle \bar{\nu}%
(p^{\prime })\gamma ^{\mu }L \tau(p) \nonumber \\
&+&(C_{3}+C_{4})\langle K\pi |\bar{s}u|0\rangle \bar{\nu}%
(p^{\prime })R \tau(p) + C_{5}\langle K\pi |\bar{s}\sigma _{\mu
\upsilon }u|0\rangle
\bar{\nu}(p^{\prime })\sigma ^{\mu \upsilon }R \tau(p)\Big] , \nonumber %
\end{eqnarray}
where $C_i$ stand for $C_i^{SUSY}$
We consider the following two interesting scenarios:\\

$i)$ The case of $C_{3}$ or $C_{4}$ gives relevant
contributions while $C_5$ is negligible. In this case,
SUSY induces a relative weak phase between the
vector and scalar form factors describing
this process.\\

$ii)$ The case of $C_{5}$ gives relevant
contributions while $C_{3,4}$ are negligible. In this
case, SUSY induces a relative
weak phase between the vector and tensor form factors. A CP asymmetry in decay rate could be measured.\\

\subsection{First scenario}
\begin{equation}
\langle K\pi |\bar{s}\gamma _{\mu }u|0\rangle =f_{V}(t)Q_{\mu
}+f_{S}(t)(q+q^{\prime })_{\mu }~,\Rightarrow \langle K\pi
|\bar{s}u|0\rangle =\frac{t}{m_{s}-m_{u}}f_{S}(t)~,\nonumber
\end{equation}

\begin{eqnarray}
\mathcal{A}_{T}(\tau \rightarrow K\pi \nu ) &=&\frac{G_{F}V_{us}}{\sqrt{%
2}}(1+C_{1})\times   \nonumber \\
&&\!\!\!\!\!\!\!\!\left\{ f_{V}Q^{\mu }\bar{u}(p^{\prime })\gamma
_{\mu
}Lu(p)+\left[ m_{\tau }+\left( \frac{C_{3}+C_{4}}{1+C_{1}}\right) \frac{t}{%
m_{s}-m_{u}}\right] f_{S}\bar{u}(p^{\prime })Ru(p)\right\} \
.\nonumber
\end{eqnarray}
Using CLEO limit, we can translate this bound into:
\begin{equation} -0.010\leq Im\left(
\frac{C_{3}+C_{4}}{1+C_{1}}\right) \leq 0.004\ ,\nonumber
\end{equation}%
where we have used $m_{s}-m_{u}=100$ MeV, and the
average value $\langle t\rangle \approx (1332.8\ \mbox{\rm
MeV})^{2}$. Using $M_{1}=100$ and $M_{2}=200$ GeV
\ and $\mu =M_{\tilde{q}}=400$ GeV and $\tan \beta =20$, one gets
\begin{equation} Im\left(
\frac{C_{3}+C_{4}}{1+C_{1}}\right) \simeq 1.3\times 10^{-5}Im(\delta
_{21}^{d})_{RL} \nonumber
\end{equation}
\subsection{Second scenario}
\begin{eqnarray} \mathcal{A}_{T}(\tau
\rightarrow K\pi \nu ) &=& \frac{G_{F}V_{us}}{\sqrt{2}}(1+C_{1})
\left\{ f_{V}(t)Q_{\mu }\bar{u}%
(p^{\prime })\gamma ^{\mu }Lu(p)\right.  \nonumber \\
&+&\left. \frac{C_{5}}{1+C_{1}}\langle K\pi |\bar{s}\sigma _{\mu
\upsilon }u|0\rangle \bar{u}(p^{\prime })\sigma ^{\mu \upsilon
}Ru(p)\right\} \nonumber
\end{eqnarray}
\begin{equation}
\langle K\pi |\bar{s}\sigma _{\mu \upsilon }u|0\rangle =\frac{ia}{m_{K}}%
\left[ (p_{\pi })^{\mu }(p_{K})^{\nu }-(p_{\pi })^{\nu }(p_{K})^{\mu
}\right] \nonumber
\end{equation}
where $a$ is a dimensionless quantity which fixes
the scale of the hadronic matrix element.
\begin{eqnarray}
a_{CP} &=&\frac{\Gamma (\tau ^{-}\rightarrow K^{-}\pi ^{0}\nu _{\tau
})-\Gamma (\tau ^{+}\rightarrow K^{+}\pi ^{0}\nu _{\tau })}{\Gamma
(\tau ^{-}\rightarrow K^{-}\pi ^{0}\nu _{\tau })+\Gamma (\tau
^{+}\rightarrow K^{+}\pi ^{0}\nu _{\tau })}
 \simeq \frac{a}{2}Im \ C_{5} \nonumber \\
&\simeq &1.4\times 10^{-7}a \ Im (\delta _{21}^{u})_{LR} \nonumber
\end{eqnarray}

\section{Conclusion}
The goal of this work is to illustrate the difficulties to define an observable CP asymmetry in semi-leptonic $\tau$ decays. We show that in standard model, a CP asymmetry in the total decay rate can be induced through higher order contributions but the absolute value of this CP asymmety is too small to be once accessible to experiments. In the supersymmetric case, we have computed the effective hamiltonian derived
from SUSY for $|\Delta S|=1$ tau lepton decays using the mass
insertion approximation. Supersymmetric extensions of the SM could induce
CP violating asymmetry in the double differential
distribution as CLEO collaboration did  and could also induce
CP asymmetry in total decay rate due to
interference between $O_{5}$ and $O_{1}$
operators. A direct consequence of our computation is that any CP
asymmetry in the channel under consideration bigger
than $10^{-6}$ will be a clear evidence of not only
Physics beyond Standard Model but also an
evidence of Physics beyond SUSY extensions of the SM.

\section{Acknowledgements}
D.D. wants to thank Gabriel Lopez Castro, Shaaban  Khalil, Gaber Faisel and L.T. Lopez Lozano for their collaborations in the realization of this work.
This work was financially supported by PROMEP PTC project, Conacyt Project numero 46195.

\section{Appendix}

\appendix

\section*{\textbf{ {SUSY contributions to Wilson coefficients
}}}

Here we provide the complete expressions for the supersymmeric
contributions, at leading order in MIA, for the Wilson
coefficients of $\tau
^{-}\rightarrow s\bar{u}\nu _{\tau }$ transition, $C_{i}(M_{W})$, $i=1,..,5$%
. The dominant SUSY contributions are given by
chargino-neutralino box diagram exchanges, as illustrated in Fig. 2.

The effective Hamiltonian $H_{eff}$ derived from SUSY can be expressed as
\begin{eqnarray}
H_{eff} &=&\frac{G_{F}}{\sqrt{2}}V_{us}\sum_{i}C_{i}(\mu )Q_{i}(\mu ), \\
&=&\sum_{i}\tilde{C}_{i}(\mu )Q_{i}(\mu ),
\end{eqnarray}%
where $C_{i}$ are the dimensionless Wilson coefficients and
$Q_{i}$ are the relevant local operators at low energy scale $\mu
\simeq m_{\tau }$.
In terms of the vertex, one can write the complete vertex as a product of
the vertex coming from leptonic sector and of the vertex coming from
hadronic sector. In this respect we can also write the Wilson coefficients
as
\begin{eqnarray*}
\tilde{C}_{i} &=&C_{i(\tau -\chi ^{-})}^{l}\left( C_{i(\tau -\chi
^{-}-s)}^{q}+C_{i((\tau -\chi ^{-}-u)}^{q}\right) \\
&&+C_{i(\tau -\chi ^{0})}^{l}\left( C_{i(\tau -\chi ^{0}-s)}^{q}+C_{i((\tau
-\chi ^{0}-u)}^{q}\right)
\end{eqnarray*}%
where the $C_{i}^{l}$ is due to the leptonic vertex and $C_{i}^{q}$ is from
the quark sectors. If we expand $C_{i}^{l,q}$ in terms of the mass
insertions, one finds that the leading contributions are given by

\begin{eqnarray}
\tilde{C}_{3} &=&C_{3(\tau -\chi ^{-})}^{l(0)}C_{3(\tau -\chi
^{-}-s)}^{q(1)}I_{n}(x_{i},x_{j})+C_{3(\tau -\chi
^{0})}^{l(0)}C_{3(\tau -\chi ^{0}-s)}^{q(1)}I_{n}(x_{i},x_{j})
\nonumber\\
&&+O(\delta ^{2}),
\end{eqnarray}
where $I_n(x_i,x_j)$ is defined below and
$x_i=m_{\chi_i^{\pm}}^2/\tilde{m}^2$ and
$x_j=m_{\chi_j^{0}}^2/\tilde{m}^2$.
\begin{eqnarray}
C_{3(\tau -\chi ^{0})}^{l(0)} &=&g(h_{e})_{33}N_{i3}U_{j1}^{\ast
}(U_{MNS}^{\ast })_{33} \\
&&-g\sqrt{2}\tan \theta _{w}N_{i1}U_{j2}^{\ast
}(h_{e})_{33}(U_{MNS}^{\ast })_{33},
\end{eqnarray}%
\begin{equation} C_{3(\tau -\chi
^{-})}^{l(0)}=-(h_{e})_{33}\frac{g}{\sqrt{2}}(N_{i2}-\tan \theta
_{w}N_{i1})U_{j2}(U_{MNS}^{\ast })_{33}, \end{equation}%

\begin{eqnarray}
C_{3(\tau -\chi ^{-}-s)}^{q(1)} &=&(-\frac{1}{8})\left( \frac{g^{2}}{\sqrt{2}}\frac{2}{3}%
\tan \theta _{w}N_{i1}^{\ast }U_{j1}^{\ast }(V_{CKM}^{\ast
})_{1a}(\delta
_{RL}^{d})_{2a}\right.\nonumber \\
&&\left. -\frac{g}{\sqrt{2}}\frac{2}{3}\tan \theta
_{w}N_{i1}^{\ast }U_{j2}^{\ast }(h_{d})_{33}(V_{CKM}^{\ast
})_{13}(\delta _{RR}^{d})_{23}\right),
\end{eqnarray}

\begin{eqnarray}
C_{3(\tau -\chi ^{0}-s)}^{q(1)} &=&(-\frac{1}{8})\left( \frac{2}{3}\frac{g^{2}}{\sqrt{2}}%
\tan \theta _{w}U_{j1}^{\ast }N_{i1}^{\ast }(V_{CKM}^{\ast
})_{1a}(\delta
_{RL}^{d})_{2a}\right.\nonumber \\
&&\left. -\frac{2}{3}\frac{g}{\sqrt{2}}\tan \theta
_{w}U_{j2}^{\ast} N_{i1}^{\ast }(V_{CKM}^{\ast })_{13}(
\delta_{RR}^{d})_{23}(h_{d})_{33}\right).
\end{eqnarray}

\begin{equation} \tilde{C}_{4}=C_{4(\tau -\chi ^{-})}^{l(0)}C_{4(\tau -\chi
^{-}-u)}^{q(1)}\tilde{I}_{n}(x_{i},x_{j})+C_{4(\tau -\chi
^{0})}^{l(0)}C_{4((\tau -\chi
^{0}-u)}^{q(1)}\tilde{I}_{n}(x_{i},x_{j})+O(\delta ^{2}),
 \end{equation}
where $\tilde{I}_{n}(x_{i},x_{j})$ is given below.

\begin{eqnarray}
C_{4(\tau -\chi ^{-})}^{l(0)}
&=&-(h_{e})_{33}\frac{g}{\sqrt{2}}(N_{i2}-\tan
\theta _{w}N_{i1})U_{j2}(U_{MNS}^{\ast })_{33} \nonumber\\
&=&C_{3(\tau -\chi ^{-})}^{l(0)},
\end{eqnarray}

\begin{eqnarray}
C_{4(\tau -\chi ^{0})}^{l(0)} &=&g(h_{e})_{33}N_{i3}U_{j1}^{\ast
}(U_{MNS}^{\ast
})_{33} \nonumber\\
&&-g\sqrt{2}\tan \theta _{w}N_{i1}U_{j2}^{\ast
}(h_{e})_{33}(U_{MNS}^{\ast })_{33},
 \end{eqnarray}

\begin{eqnarray}
C_{4(\tau -\chi ^{0}-u)}^{q(1)} &=&(-\frac{1}{8})\left( -\frac{4}{3}\frac{g^{2}}{\sqrt{2}}%
\tan \theta _{w}N_{i1}V_{j1}(V_{CKM}^{\ast })_{a2}(\delta
_{LR}^{u})_{a1}\right.  \nonumber\\
&&\left. +\frac{4}{3}\frac{g}{\sqrt{2}}\tan \theta
_{w}N_{i1}V_{j2}(h_{u})_{33}(V_{CKM}^{\ast })_{32}(\delta
_{RR}^{u})_{31}\right),
\end{eqnarray}%

\begin{eqnarray}
C_{4(\tau -\chi ^{-}-u)}^{q(1)} &=&(-\frac{1}{8})\left( -\frac{4}{3}\frac{g^{2}}{\sqrt{2}}%
\tan \theta _{w}V_{j1}N_{i1}(V_{CKM}^{\ast })_{a2}(\delta
_{LR}^{u})_{a1}\right.  \nonumber\\
&&\left. +\frac{4}{3}\frac{g}{\sqrt{2}}\tan \theta
_{w}V_{j2}N_{i1}(h_{u})_{33}(V_{CKM}^{\ast })_{32}(\delta
_{RR}^{u})_{31}\right).
\end{eqnarray}

 \begin{equation}
\tilde{C}_{5(\tau -\chi ^{0}-u)}=-\frac{1}{4}\tilde{C}_{4(\tau
-\chi ^{0}-u)}, \end{equation}

\begin{equation}
\tilde{C}_{5(\tau -\chi ^{-}-u)}=\frac{1}{4}\tilde{C}_{4(\tau
-\chi ^{-}-u)}. \end{equation}

The loop integrals $I_{n}(x_{i},x_{j})$ and
$\tilde{I}_{n}(x_{i},x_{j})$ are defined as follows:

\begin{eqnarray}
I(x_{i},x_{j}) &=&\frac{1}{16\pi ^{2}\tilde{m}^{2}}\left( \frac{1}{%
x_{i}-x_{j}}\right) \left( \frac{x_{i}^{2}-x_{i}-x_{i}^{2}logx_{i}}{%
(1-x_{i})^{2}}-(x_{i}\leftrightarrow x_{j})\right) ,  \nonumber \\
\widetilde{I}(x_{i},x_{j}) &=&\frac{\sqrt{x_{i}x_{j}}}{16\pi ^{2}\tilde{m}%
^{2}}\left( \frac{1}{x_{i}-x_{j}}\right) \left( \frac{%
x_{i}^{2}-x_{i}-x_{i}logx_{i}}{(1-x_{i})^{2}}-(x_{i}\leftrightarrow
x_{j})\right) ,  \nonumber \\
I_{n}(x_{i},x_{j}) &=&\frac{1}{32\pi ^{2}\tilde{m}^{2}}\left( \frac{1}{%
x_{i}-x_{j}}\right) \left( \frac{2x_{i}^{2}-2x_{i}-2x_{i}logx_{i}}{%
(x_{i}-1)^{2}}-\frac{x_{i}^{3}-4x_{i}^{2}+3x_{i}+2x_{i}logx_{i}}{%
(x_{i}-1)^{3}}\right.\nonumber\\
&-&\left.(x_{i}\leftrightarrow x_{j})\right) ,~~~  \nonumber \\
\tilde{I_{n}}(x_{i},x_{j}) &=&\frac{-\sqrt{x_{i}x_{j}}}{32\pi ^{2}\tilde{m}%
^{2}}\left( \frac{1}{x_{i}-x_{j}}\right) \left( \frac{%
x_{i}^{3}-4x_{i}^{2}+3x_{i}+2x_{i}logx_{i}}{(x_{i}-1)^{3}}%
-(x_{i}\leftrightarrow x_{j})\right) .
\end{eqnarray}
The expression for the other Wilson coefficients can be found in ref.\cite{david2}.



\end{document}